\documentclass[11pt,aps,preprint,amsmath,amssymb,t1]{revtex4-1}
\usepackage[T1]{fontenc}
\usepackage[latin9]{inputenc}
\usepackage{amstext}
\usepackage{graphicx}

\makeatletter

\providecommand{\tabularnewline}{\\}

\usepackage{epstopdf}
\usepackage{slashed}

\makeatother

\begin{document}
\title{Study on Anomalous Neutral Triple Gauge Boson Couplings from Dimension-eight
Operators at the HL-LHC}
\author{A. Senol}
\email[]{senol_a@ibu.edu.tr}

\author{H. Denizli}
\email[]{denizli_h@ibu.edu.tr}

\affiliation{Department of Physics, Bolu Abant Izzet Baysal University, 14280,
Bolu, Turkey}
\author{A. Yilmaz}
\email[]{aliyilmaz@giresun.edu.tr}

\affiliation{Department of Electrical and Electronics Engineering, Giresun University,
28200 Giresun, Turkey}
\author{I. Turk Cakir}
\email[]{ilkay.turk.cakir@cern.ch}

\affiliation{Department of Energy Systems Engineering, Giresun University, 28200
Giresun, Turkey}
\author{O. Cakir}
\email[]{ocakir@science.ankara.edu.tr}

\affiliation{Department of Physics, Ankara University, 06100, Ankara, Turkey}
\date{\today}
\begin{abstract}
The anomalous neutral triple gauge boson couplings (aNTGCs) for the
$Z\gamma\gamma$ and $Z\gamma Z$ vertices described by dimension-eight
operators are examined through the process $pp\to l^{+}l^{-}\gamma$
at the High-Luminosity Large Hadron Collider (HL-LHC). We performed
an analysis on transverse momentum of photon and angular distribution
of charged lepton in the final state including detector effects. Sensitivity
limits of the $C_{\widetilde{B}W}$, $C_{BB}$ couplings are obtained
at $95\%$ C.L. to constrain for the range $[-1.88;1.88]$ TeV$^{-4}$,
$[-1.47:1.47]$ TeV$^{-4}$ and $[-1.14:1.14]$ TeV$^{-4}$, $[-0.86;0.86]$
TeV$^{-4}$ with an integrated luminosity of $300$ fb$^{-1}$ and
$3000$ fb$^{-1}$, respectively.
\end{abstract}
\maketitle

\section{Introduction}

The Standard Model (SM) through the non-Abelian $SU(2)_{L}\times U(1)_{Y}$
gauge group of the electroweak sector predicts the gauge boson self-interactions.
The tree-level vertices of three neutral gauge bosons are not allowed
since it violates the underlying $SU(2)_{L}\times U(1)_{Y}$ symmetry.
Deviations of triple gauge couplings from SM expectations can provide
clues about the new physics beyond the SM. The effect of new physics
can be parametirized in a model independent way by an effective Lagrangian
at TeV energy scale. Using effective field theory (EFT), the Lagrangian
for neutral triple gauge boson interactions can be written as \citep{Degrande:2013kka}
\begin{eqnarray}
\mathcal{L}^{nTGC}=\mathcal{L}^{SM}+\sum_{i}\frac{C_{i}}{\Lambda^{4}}(\mathcal{O}_{i}+\mathcal{O}_{i}^{\dagger})\label{eq:1}
\end{eqnarray}
where $\Lambda$ is the new physics scale, $i$ runs over the label
of the four operators expressed as 
\begin{eqnarray}
\mathcal{O}_{BW} & = & iH^{\dagger}B_{\mu\nu}W^{\mu\rho}\{D_{\rho},D^{\nu}\}H\label{eq:2}\\
\mathcal{O}_{WW} & = & iH^{\dagger}W_{\mu\nu}W^{\mu\rho}\{D_{\rho},D^{\nu}\}H\label{eq:3}\\
\mathcal{O}_{BB} & = & iH^{\dagger}B_{\mu\nu}B^{\mu\rho}\{D_{\rho},D^{\nu}\}H\label{eq:4}\\
\mathcal{O}_{\tilde{B}W} & = & iH^{\dagger}\tilde{B}_{\mu\nu}W^{\mu\rho}\{D_{\rho},D^{\nu}\}H\label{eq:5}
\end{eqnarray}
where $\tilde{B}_{\mu\nu}$ is the $B$-field strength tensor.

The coefficients $C_{\widetilde{B}W}$ (CP conserving) and $C_{BB}$,
$C_{BW}$ , $C_{WW}$ (CP violating) of dimension-eight operators
describe anomalous Neutral Triple Gauge Couplings (aNTGC). The contributions
from new physics is expected to be suppressed by the inverse powers
of the scale of new physics. When the new physics appears at high
energy scale, the largest contribution to the subprocess $q\bar{q}\to Z\gamma$
is expected from the interference between the SM and the dimension-eight
operators. The resulting matrix-element squared for the process $pp\to Z\gamma$
(where $Z\to l^{+}l^{-}$ with $l^{\pm}=e^{\pm},\mu^{\pm}$) is given
by

\begin{equation}
|M|^{2}=|M_{SM}|^{2}+2Re(M_{SM}M_{D8}^{*})+|M_{D8}|^{2}.\label{eq:6}
\end{equation}
Here, the last term could be small due to the factor $\sim C_{i}^{2}\Lambda^{-8}$
when the $\Lambda$ kept as high energy scale. However, second term
may contribute importantly since the interference takes contribution
proportional to $\sim C_{i}\Lambda^{-4}$. The total cross section
for $pp\to l^{+}l^{-}\gamma$ process can be written as $\sigma_{tot}=\sigma_{SM}+\sigma_{D8}+\sigma_{Int}$
with the leading order SM cross section $\sigma_{SM}$, the dimension-eight
term cross section $\sigma_{D8}$ and the interference term cross
section $\sigma_{Int}$.

Although the dimension-six operators do not induce aNTGC at the tree-level,
they can have an effect at the one-loop level. The one-loop contributions
from dimension-six operators would be of the order $(\alpha/4\pi)(\hat{s}/\Lambda^{2})$
while the tree-level contribution from the dimension-eight operators
are of the order $(\hat{s}v^{2}/\Lambda^{4})$ \citep{Degrande:2013kka}.
As a result, the contribution of the dimension-eight operators dominates
the one-loop contribution of the dimension-six operators for $\Lambda\lesssim2v\sqrt{\pi/\alpha}\approx10$
TeV.

We have studied anomalous neutral triple gauge boson couplings from
dimension-eight operators via the $pp\to l^{+}l^{-}\gamma$ process
at High Luminosity Large Hadron Collider (HL-LHC) with 14 TeV center
of mass energy, and we expect an enhancement due to the existence
of aNTGCs with high $p_{T}$ photon in the final state \citep{Barger:1984yn,Baur:1992cd,Baur:1997kz}.
The HL-LHC may provide a portal to complete opportunities at the LHC
for discovery of new physics beyond the SM. The HL-LHC program as
a top priority in particle physics in the context of developing the
strategy for particle physics \citep{ATLAS-Collaboration:2012jwa,ATLAS-collaboration:2012iza}
is planned at two benchmark values of integrated luminosity: the 300
fb$^{-1}$ expected by the end of Run 3, and the 3000 fb$^{-1}$ expected
to be delivered by the HL-LHC \citep{ATLAS:2013hta}.

\section{Cross Sections}

\begin{figure}[!hbt]
\centering \includegraphics[width=0.95\textwidth,height=0.4\textheight]{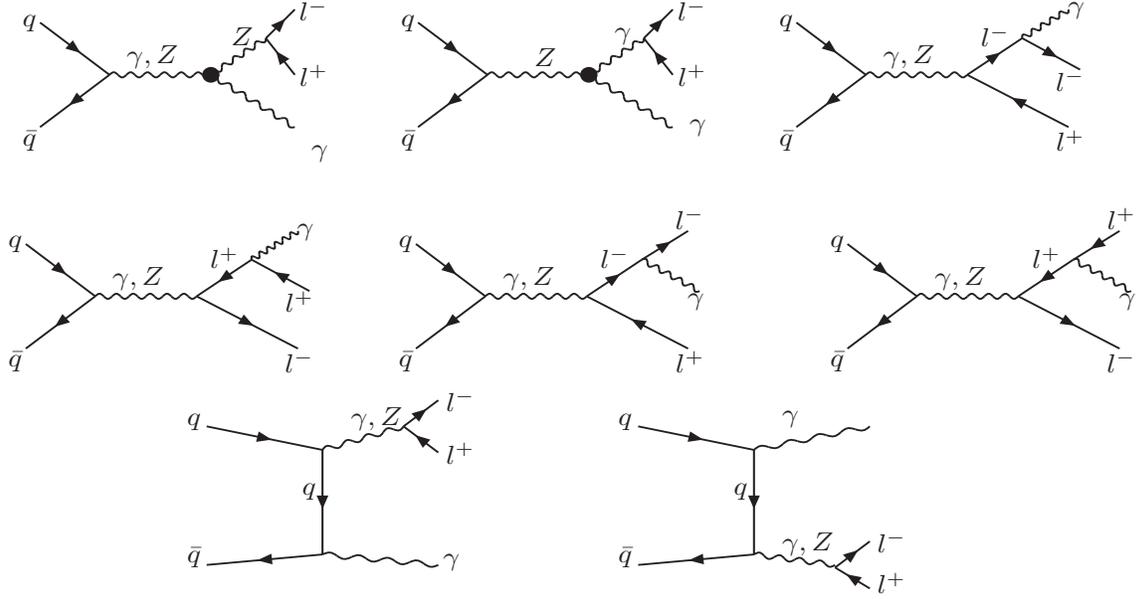}
\caption{Feynman diagrams for subprocess $q\bar{q}\rightarrow l^{-}l^{+}\gamma$
contributed from anomalous $Z\gamma\gamma$ and $Z\gamma Z$ vertices
(first two) and the SM. \label{fig:fig1}}
\end{figure}

\begin{figure}[!hbt]
\centering \includegraphics[scale=0.8]{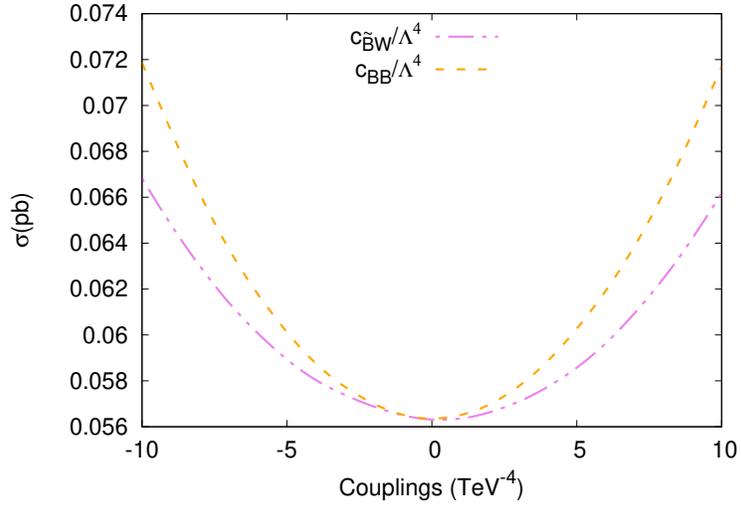} \caption{The signal cross sections for process $pp\rightarrow l^{-}l^{+}\gamma$
, with photon transverse momentum $p_{T}^{\gamma}>100$ GeV, depending
on dimension-eight couplings at HL-LHC. \label{fig:fig2}}
\end{figure}

The leading order Feynman diagrams corresponding to the process $pp\to l^{+}l^{-}\gamma$
are given in Fig.~\ref{fig:fig1}. In this figure, the first diagram
contains the anomalous $Z\gamma\gamma$ and $ZZ\gamma$ couplings,
and second diagram contains only the anomalous $Z\gamma\gamma$ couplings,
while the others come from SM electroweak processes. The operators
in Eqs. (2)-(5) are implemented through FeynRules package \citep{Alloul:2013bka}
as a Universal FeynRules Output (UFO) \citep{Degrande:2011ua}. After
implementation of this UFO model file the cross section of $pp\to l^{+}l^{-}\gamma$
process at the center of mass energy of 14 TeV is calculated with
\verb|MadGraph5_aMC@NLO| \citep{Alwall:2014hca}. In the study, we
focus on CP-even $C_{\widetilde{B}W}$ coupling and CP-odd $C_{BB}$
coupling because the deviation in cross section from its SM value
for these couplings is larger than that for $C_{BW}$, $C_{WW}$ as
mentioned in Ref. \citep{Senol:2018cks}. The Fig.~\ref{fig:fig2}
shows the cross sections of the $pp\to l^{+}l^{-}\gamma$ process
as a function of two dimension-eight couplings $C_{\widetilde{B}W}$,
$C_{BB}$. The cross sections are calculated via generator level cuts
as follows:
\begin{itemize}
\item photon transverse momentum $p_{T}^{\gamma}>100$ GeV
\item photon pseudorapidity $|\eta^{\gamma}|<2.5$
\item charged lepton transverse momentum $p_{T}^{l}>10$ GeV and pseudorapidity
$|\eta^{l}|<2.5$
\item the invariant mass of final state charged leptons $m_{ll}>50$ GeV
\item charged lepton - photon separation $\Delta R(l,\gamma)>0.7$, the
separation between the charged lepton and photon in the pseudorapidity-azimuthal
angle plane is defined by
\end{itemize}
\begin{equation}
\Delta R(l,\gamma)=\left[(\Delta\phi_{l,\gamma})^{2}+(\Delta\eta_{l,\gamma})^{2}\right]^{1/2}.\label{eq:7}
\end{equation}
For the calculation of cross sections, only one coupling at a time
is varied from its SM value.

\section{analysis framework}

\begin{figure}
\includegraphics[scale=0.35]{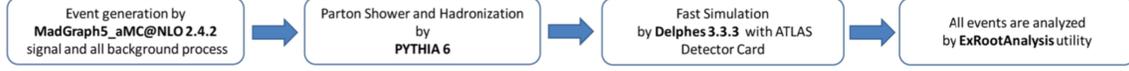} \caption{A brief view of analysis flowchart chain. \label{fig:fig3}}
\end{figure}

The study on effective dimension-eight aNTG couplings $C_{\tilde{B}W}$,
$C_{BB}$ and SM contribution as well as interference between effective
couplings and SM contributions have been performed via $pp\to l^{+}l^{-}\gamma$
process where $l^{\pm}=e^{\pm},\mu^{\pm}$. For the detailed analysis
we follow steps as shown in Fig~\ref{fig:fig3}. The signal and background
events are generated with \verb|MadGraph5_aMC@NLO| applying generator-level
cuts for pseudo-rapidity $|\eta^{l,\gamma}|<2.5$, and transverse
momentum $p_{T}^{l,\gamma}>20$ GeV and passed through the Pythia
6 \citep{Sjostrand:2006za} for parton showering and hadronization.
The detector effects are included by ATLAS detector card in \verb|Delphes 3.3.3|
\citep{deFavereau:2013fsa} package. Then, all events are analyzed
with with ROOT \citep{Brun:1997pa} by using the ExRootAnalysis utility
\citep{exroot}.

\begin{figure}[!hbt]
\includegraphics[scale=0.4]{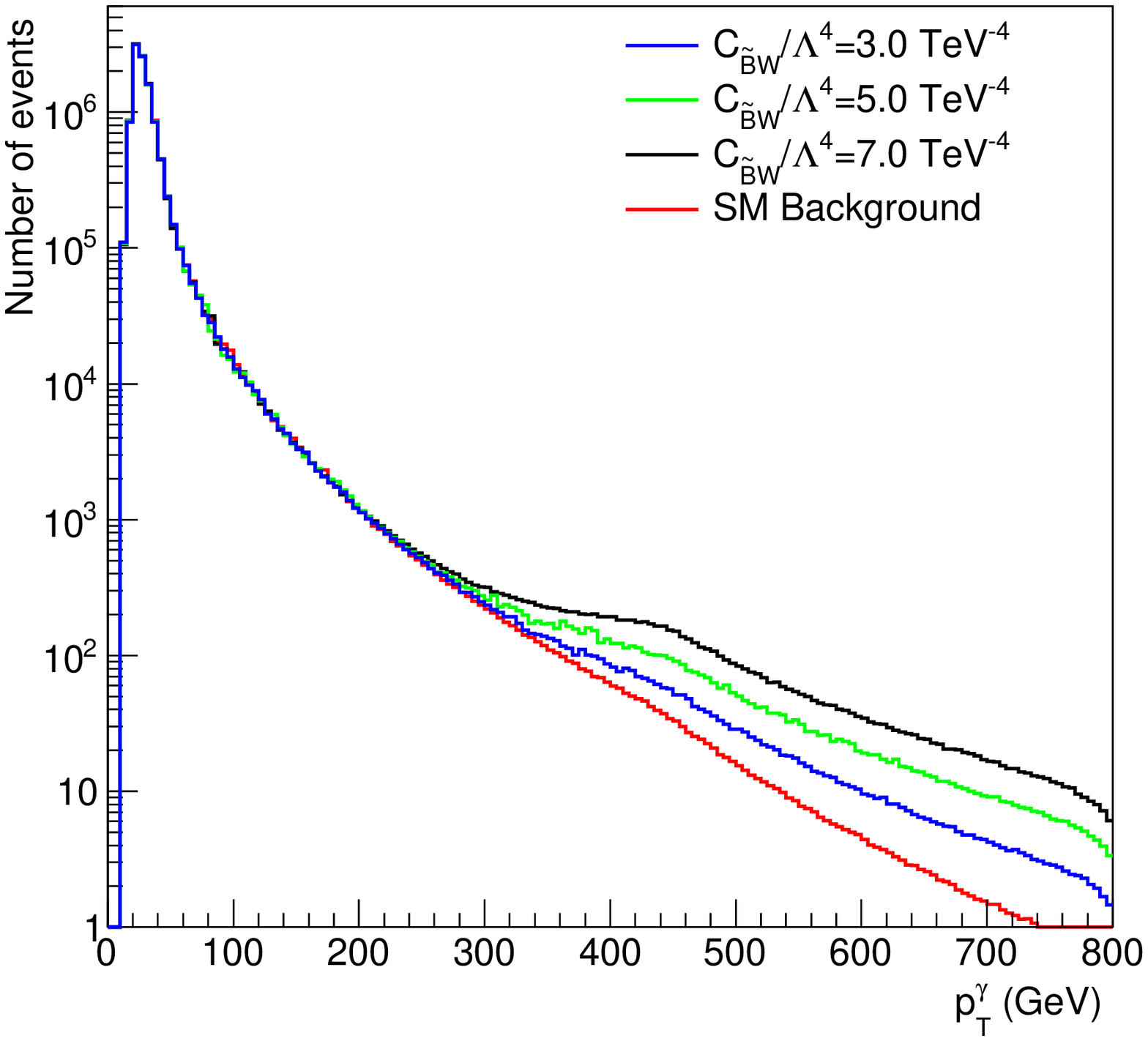} \includegraphics[scale=0.4]{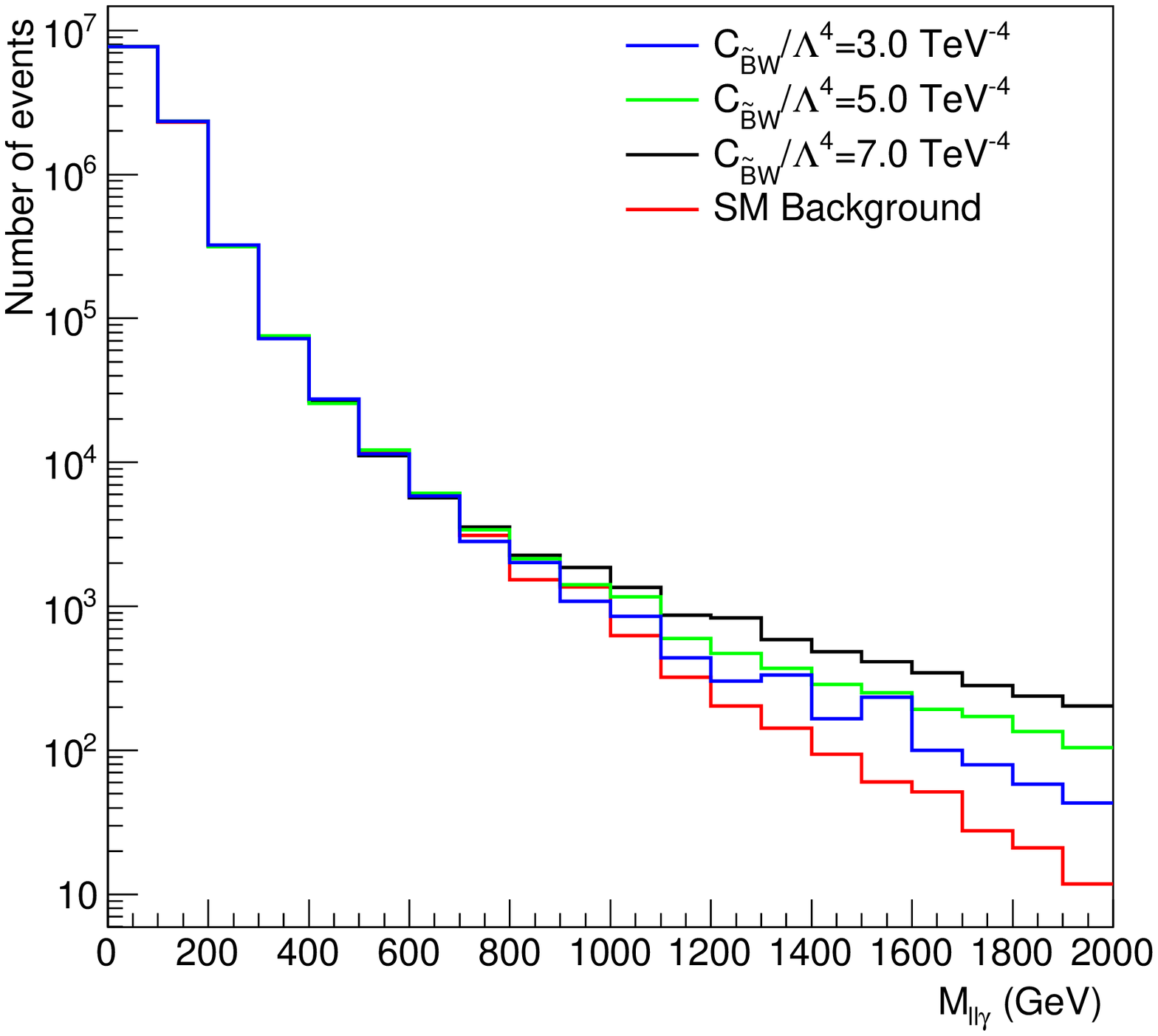}
\caption{The transverse momentum distribution of photon $p_{T}^{\gamma}$ (left),
and the invariant mass distribution $M_{ll\gamma}$ (right), for three
different values of coupling $C_{\widetilde{B}W}/\Lambda^{4}$ and
SM background of $pp\rightarrow l^{-}l^{+}\gamma$ process. \label{fig:fig4}}
\end{figure}

\begin{figure}[!hbt]
\includegraphics[scale=0.4]{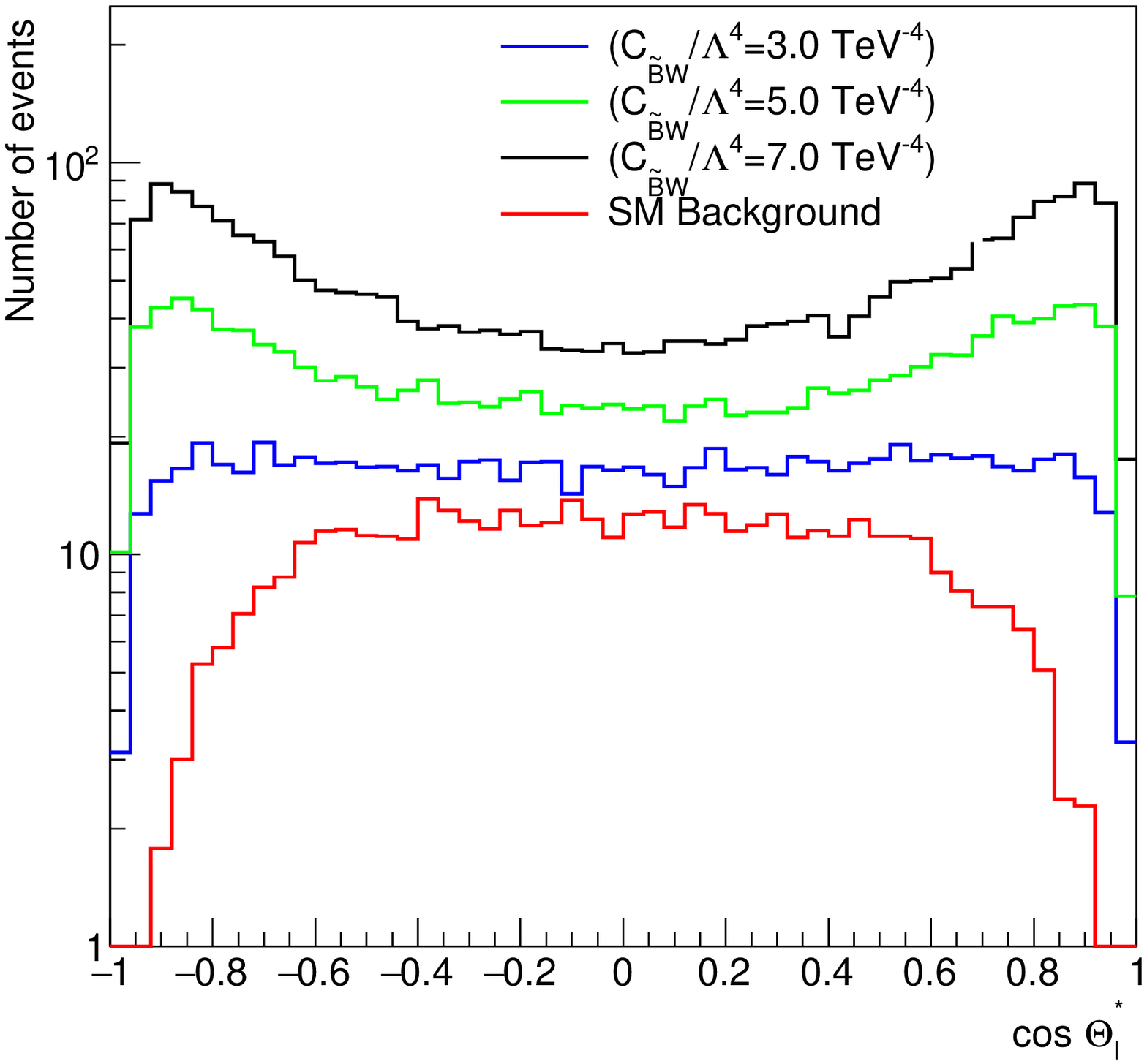} \includegraphics[scale=0.4]{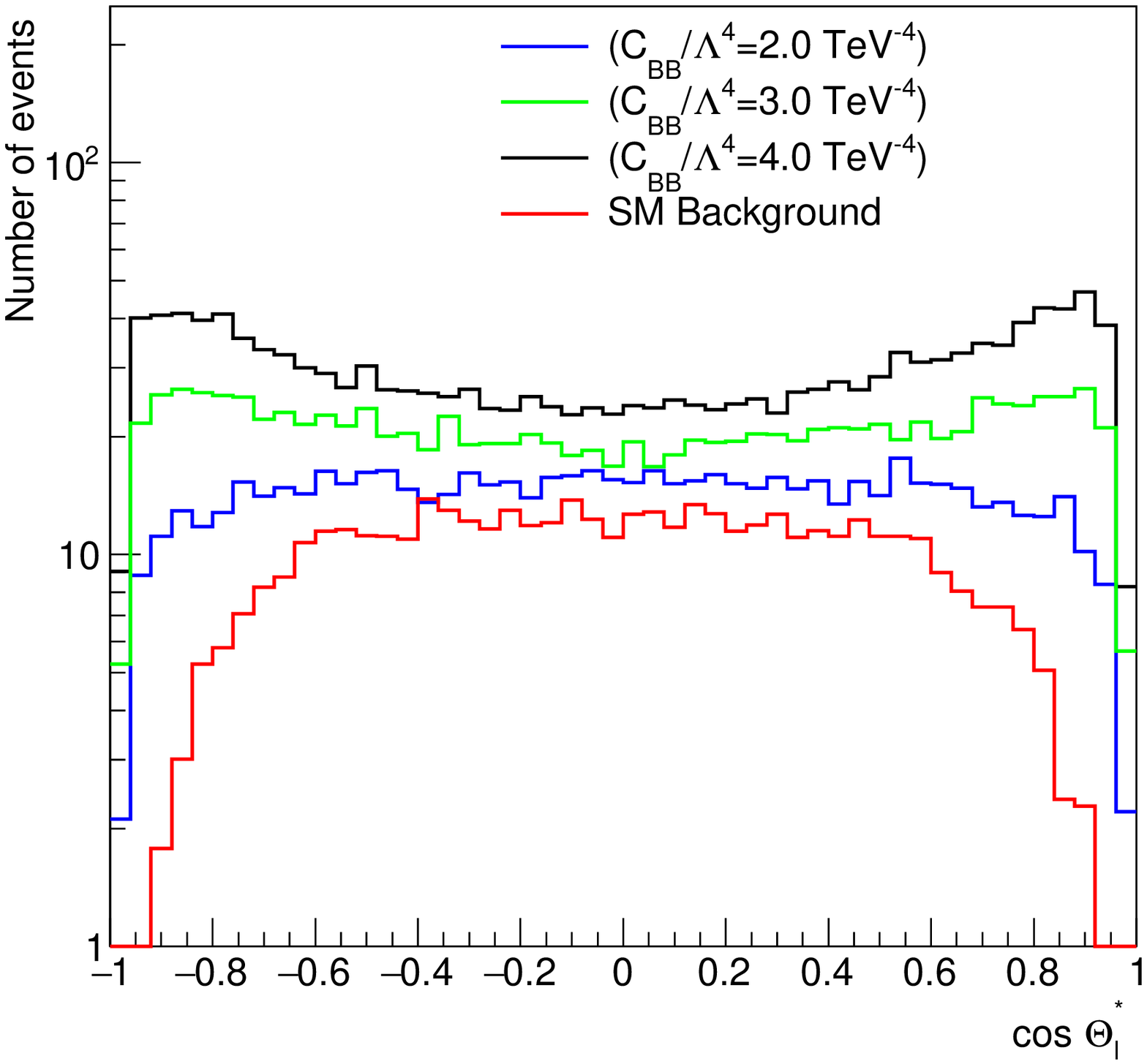}
\caption{The $\cos\theta_{l}^{*}$ distributions for $C_{\widetilde{B}W}/\Lambda^{4}$
(left panel) and $C_{BB}/\Lambda^{4}$ (right panel) and SM background
of the $pp\to l^{-}l^{+}\gamma$ process. \label{fig:fig5}}
\end{figure}

For the event selection, we require one photon and at least two charged
leptons ($l^{\pm}=e^{\pm},\mu^{\pm}$); same flavor but opposite sign
and the angular separation between lepton and photon $\Delta R(l,\gamma)>0.7$.
As seen from Fig.~\ref{fig:fig4}(left pad), the transverse momentum
distribution of photon (in the final state for $pp\to l^{+}l^{-}\gamma$)
for the signal deviates significantly from that of SM background for
all couplings starting from $200$ GeV. The invariant mass distributions
of the $l^{+}l^{-}\gamma$ system for signal are shown in Fig.~\ref{fig:fig4}(right
pad). The deviations at large values of $m_{ll\gamma}>500$ GeV become
more pronounced. Therefore, we impose the following cuts in addition
to above mentioned initial cuts: (a) $p_{T}^{\gamma}>400$ $(300)$
GeV, (b) $m_{ll\gamma}>500$ GeV and (c) $m_{ll}>50$ GeV in order
to separate signal from the SM background efficiently. Fig.~\ref{fig:fig5}
shows $\cos\theta_{l}^{*}$ distributions of signal for $C_{\widetilde{B}W}/\Lambda^{4}$
(left pad), $C_{BB}/\Lambda^{4}$ (right pad) couplings and SM background.
Here, $\cos\theta_{l}^{*}$ is the polar angle in the $l^{+}l^{-}$
rest frame with respect to the $l^{+}l^{-}$ direction in the $l^{+}l^{-}\gamma$
rest frame. Since the angular distribution of final state particles
from signal and background processes are different, we have used this
distribution as tool to obtain attainable limits on effective dimension-eight
aNTG couplings. Distributions given in Fig.~\ref{fig:fig4} and Fig.~\ref{fig:fig5}
are normalized to the number of expected events which is defined to
be the cross section of each processes times integrated luminosity
of $L_{int}=3000$ fb$^{-1}$.

\begin{figure}[!hbt]
\includegraphics[scale=0.6]{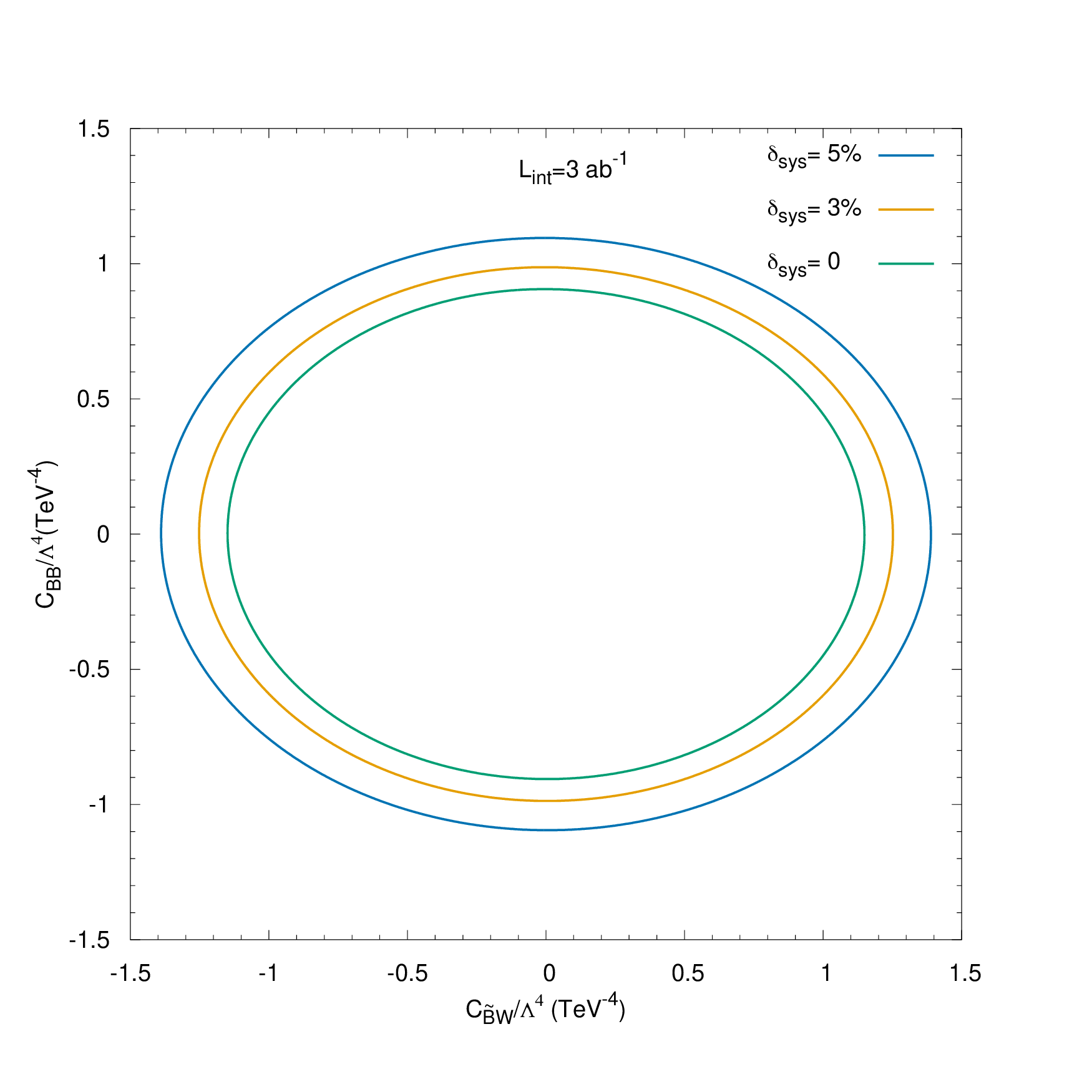} \caption{Two-dimensional $95\%$ C.L. intervals in planes of $C_{\widetilde{B}W}/\Lambda^{4}$
and $C_{BB}/\Lambda^{4}$ . \label{fig:fig6}}
\end{figure}

In order to obtain $95\%$ C.L. limits on the aNTG couplings, a $\chi^{2}$
criterion with and without systematic error is applied. Here $\chi^{2}$
function is defined as follows 
\begin{eqnarray}
\chi^{2}=\sum_{i}^{n_{bins}}\left(\frac{N_{i}^{NP}-N_{i}^{B}}{N_{i}^{B}\Delta_{i}}\right)^{2}
\end{eqnarray}
where $N_{i}^{NP}$ is the total number of events in the existence
of effective couplings, $N_{i}^{B}$ is total number of events of
the corresponding SM backgrounds in $i$th bin of the $\cos\theta_{l}^{*}$
distributions, $\Delta_{i}=\sqrt{\delta_{sys}^{2}+1/N_{i}^{B}}$ is
the combined systematic ($\delta_{sys}$) and statistical errors in
each bin. The one-parameter $\chi^{2}$ results of signal events obtained
from $cos\theta_{l}^{*}$ distributions are $C_{\widetilde{B}W}/\Lambda^{4}=3.0,5.0,7.0$
TeV$^{-4}$ and $C_{BB}/\Lambda^{4}=2.0,3.0,4.0$ TeV$^{-4}$ as given
in Tables ~\ref{tab1} and \ref{tab2}, respectively. In these tables,
a cut on the photon transverse momentum $p_{T}^{\gamma}>400$ GeV
and integrated luminosity of 3000 fb$^{-1}$ are considered while
only one coupling at a time is varied from its SM value. The two-dimensional
$95\%$ C.L. intervals in planes of $C_{\widetilde{B}W}/\Lambda^{4}$
and $C_{BB}/\Lambda^{4}$ are presented in Fig. \ref{fig:fig6}. One
can also find one-dimensional confidence intervals on the relevant
parameter axes.

\section{Discussion and Conclusion}

The effects of dimension-eight operators in $Z\gamma\gamma$ and $Z\gamma Z$
vertices are investigated via the $pp\rightarrow l^{-}l^{+}\gamma$
process. Both the final state photon transverse momentum ($p_{T}^{\gamma}$)
and polar angle ($\cos\theta_{l}^{\star}$) are considered as a discriminant
to extract bounds for $C_{\widetilde{B}W}/\Lambda^{4}$ and $C_{BB}/\Lambda^{4}$
couplings. Our expected limits on dimension-eight aNTG couplings at
$95\%$ C.L. for HL-LHC with $L_{int}=300$ fb$^{-1}$ and $3000$
fb$^{-1}$ are given in Table~\ref{tab3} as $[-1.88:1.88]$ TeV$^{-4}$
and $[-1.14;1.14]$ TeV$^{-4}$ for $C_{\widetilde{B}W}/\Lambda^{4}$,
and the limits are $[-1.47:1.47]$ TeV$^{-4}$ and $[-0.86;0.86]$
TeV$^{-4}$ for $C_{BB}/\Lambda^{4}$ (where $p_{T}^{\gamma}>400$
GeV applied), respectively. The $95\%$ C.L. current limits on dimension-eight
operators converted from coefficients of dimension-six operators for
the process $pp\to ZZ\to l^{+}l^{-}l'^{+}l'^{-}$ at $\sqrt{s}=13$
TeV and $L_{int}=36.1$ fb$^{-1}$ from ATLAS Collaboration are reported
as $-5.9~\textrm{TeV}^{-4}<C_{\widetilde{B}W}/\Lambda^{4}<5.9~\textrm{TeV}^{-4}$
and $-2.7~\textrm{TeV}^{-4}<C_{BB}/\Lambda^{4}<2.8~\textrm{TeV}^{-4}$
\citep{Aaboud:2017rwm}. Comparing with the current experimental results,
we obtain $5$ and $3$ times better sensitivity for dimension-eight
couplings $C_{\widetilde{B}W}/\Lambda^{4}$ and $C_{BB}/\Lambda^{4}$
, respectively. We conclude that the limits on aNTG couplings would
be improved from the HL-LHC results once the systematic uncertainties
are kept under control.
\begin{acknowledgments}
Authors\textquoteright{} work was partially supported by Turkish Atomic
Energy Authority (TAEK) under the project grant no. 2018TAEK(CERN)A5.H6.F2-20.
\end{acknowledgments}

\newpage{} 
\begin{table}
\caption{The number of signal events and $\chi^{2}$ results for various coupling
value of $C_{\widetilde{B}W}/\Lambda^{4}$ after applied kinematic
cuts presented in the text using $\cos\theta_{l}^{*}$ distributions
of the $pp\to l^{-}l^{+}\gamma$ process with $L_{int}=3000$ fb$^{-1}$.
\label{tab1}}

\begin{ruledtabular}
\begin{tabular}{ccccc}
$C_{\widetilde{B}W}/\Lambda^{4}$ (TeV$^{-4}$)  & Number of events  & $\chi^{2}(\delta_{sys}=0)$  & $\chi^{2}(\delta_{sys}=3\%)$  & $\chi^{2}(\delta_{sys}=5\%)$ \tabularnewline
\hline 
3.0 & 811 & 269.02 & 190.31 & 125.20\tabularnewline
5.0 & 1458 & 2171.40 & 1536.12 & 1010.53\tabularnewline
7.0 & 2464 & 8743.88 & 6185.72 & 4069.24\tabularnewline
\end{tabular}
\end{ruledtabular}

\end{table}

\begin{table}
\caption{The number of signal events and $\chi^{2}$ results for various coupling
value of $C_{BB}/\Lambda^{4}$ after applied kinematic cuts presented
in the text using $\cos\theta_{l}^{*}$ distributions of the $pp\to l^{-}l^{+}\gamma$
process with $L_{int}=3000$ fb$^{-1}$. \label{tab2}}

\begin{ruledtabular}
\begin{tabular}{ccccc}
$C_{BB}/\Lambda^{4}$ (TeV$^{-4}$)  & Number of events  & $\chi^{2}(\delta_{sys}=0)$  & $\chi^{2}(\delta_{sys}=3\%)$  & $\chi^{2}(\delta_{sys}=5\%)$ \tabularnewline
\hline 
2.0 & 695 & 120.71 & 85.40 & 394.09\tabularnewline
3.0 & 1037 & 726.09 & 513.66 & 337.91\tabularnewline
4.0 & 1464 & 2199.84 & 1556.24 & 1023.77\tabularnewline
\end{tabular}
\end{ruledtabular}

\end{table}

\begin{table}
\caption{ Obtained limits on $C_{\widetilde{B}W}/\Lambda^{4}$ and $C_{BB}/\Lambda^{4}$
at 95\% C.L. with $L_{int}=300$ fb$^{-1}$ and $3000$ fb$^{-1}$
by assuming a non-zero dimension-eight operator at a time for $pp\to l^{-}l^{+}\gamma$
process. \label{tab3}}

\begin{ruledtabular}
\begin{tabular}{l|ccc|ccc}
Couplings & \multicolumn{3}{c|}{$L_{int}=300$ fb$^{-1}$} & \multicolumn{3}{c}{$L_{int}=3000$ fb$^{-1}$}\tabularnewline
(TeV$^{-4}$) & $\delta_{sys}=0$  & $\delta_{sys}=3\%$  & $\delta_{sys}=5\%$ & $\delta_{sys}=0$  & $\delta_{sys}=3\%$  & $\delta_{sys}=5\%$\tabularnewline
\hline 
$C_{\widetilde{B}W}/\Lambda^{4}$  & $[-1.88:1.88]$ & $[-1.89:1.89]$ & $[-1.92:1.92]$ & $[-1.14:1.14]$ & $[-1.22:1.22]$ & $[-1.34:1.34]$\tabularnewline
$C_{BB}/\Lambda^{4}$  & $[-1.47:1.47]$ & $[-1.49:1.49]$ & $[-1.51:1.51]$ & $[-0.86:0.86]$ & $[-0.93:0.93]$ & $[-1.02:1.02]$\tabularnewline
\end{tabular}
\end{ruledtabular}

\end{table}

\end{document}